# Thermal properties of bended graphene nanoribbons from nonequilibrium molecular dynamics

M. D. Han, L. Miao*, J. Mei, J. J. Jiang

(Electronic of science and technology of Huazhong University of Science and Technology, 430074, China)

**Abstract**

We have studied the thermal properties of bended graphene nanoribbons (GNRs) using nonequilibrium molecular dynamics simulations. The thermal conductivity of bended GNRs shows a non-monotonous relationship with the bending angle, due to the influence of chirality and Kapitza conductance. When a constant heat flux is allowed to flow, sharp temperature jump is observed at the inside corner. On the basis of the magnitude of these jumps, we have computed the Kapitza conductance as a function of bending angles. Besides, modification of the inside corner is applied to change the ability of heat transfer at the bending place. Equations to obtain the thermal conductivity of the whole structure from the thermal conductivity of each part have been derived to guide us for GNR-interconnected circuits design.

**Keywords**：thermal conductivity, Kapitza conductance, bended graphene nanoribbons

## 1. Introduction

With the development of semiconductor technology, the feature size of very large scale integrated circuit (VLSI) is continuously decreasing[1]. Compared with the traditional materials, graphene shows outstanding electronic, magnetic and optical properties[2-7] which indicates that graphene is a promising material for nanoelectronic devices since it was obtained by experiment in 2004[8]. Moreover, extensive studies have shown that graphene has excellent thermal properties[9] both theoretically[10-13] and experimentally[14-17]. This is of particular concern to the development of electronic devices.

Graphenen nanoribbons (GNRs), which are quasi-one-dimensional graphene nanostructures with smooth edges, are more favourable in nanoelectronic devices. Numerous studies have proved that the thermal properties of GNRs depend on a number of factors, including size[18-20], chirality[21], defects[22-25], interaction with substrates[26], etc. Specifically, the thermal conductivity of GNRs increase with their length due to phonon ballistic transport, while defects, impurity, and rough edges usually reduce the thermal conductivity of GNRs. Recently, GNRs with interesting morphologies and graphene-junctions have been made[27], which makes GNRs possible for IC interconnections to improve the thermal performance of VLSI in nano-level[28, 29]. Undoubtedly, the tailored geometric shapes will affect the thermal properties of GNRs[30]. Firstly, different morphologies change the chirality of GNRs, which will affect the thermal conductivity. Secondly, the thermal boundary resistance (Kapitza resistance[31]) at the graphene-junctions also affect the thermal properties of GNRs. For example, GNRs with twin grain boundaries have been studied recently which reveals the relationship between Kapitza resistance and misorientation angles[32].

* Email: miaoling@mail.hust.edu.cn

Focusing on the geometric shapes, we compute the thermal properties of bended GNRs with different angles, based on reverse non-equilibrium molecular dynamics (RNEMD). The thermal conductivity and Kapitza conductance are computed to investigate the influence of bending angle. Besides, the inside corner of these GNRs are modified to improve their thermal properties. Finally, we describe the relationship between the thermal conductivity of the whole bended structure and that of each part.

## 2. Method and model

In this work we perform RNEMD[33] using LAMMPS MD package[34] to compute the thermal conductivity of bended GNRs. Tersoff potential[35] is applied to define the carbon-carbon interactions. The MD simulation is carried at 300 K, with a timestep of 0.2 fs and a total simulation time of 0.6 ns.

Figure 1(a) shows the schematic picture for heat flux setting. The sample of GNR is divided into 50 slabs along X direction, where the first and last slabs are assigned to be the cold region while the 26th is the hot one. Exchanging the velocities of the hottest atom in the cold slab with the coldest atom in the hot slab every 80 MD steps under NVE esemble will induce a heat flux which is given by

$$J = \frac{1}{2tA_{yz}} \sum_{transfer} \frac{m}{2}(v_{hot}^2 - v_{cold}^2) \quad (1)$$

where J is the heat flux, m is the mas of the atoms, $v_{hot}$ and $v_{cold}$ are the velocities of the hottest atom from the cold region and the coldest atom of the hot region, t is the total simulation time, $A_{yz}$ is the cross sectional area defined by the width times the thickness of the GNR. The thickness of GNR is assumed to be 0.142 nm[18, 36], which is the length of the carbon–carbon bond. The heat flux imposed on the structure will form a temperature gradient, and the thermal conductivity can be obtained from Fourier law:

$$\kappa = \frac{J}{\partial T / \partial x} \qquad (2)$$

For bended GNRs, the bending angles are chosen as 30°, 60°, 90° and 120° to ensure the smooth edge, and the chirality of each part may be different. For example, bended GNRs with an angle of 30° (Figure 1(b)) or 90° (Figure 1(d)) have zigzag edge in part1, but the chirality changes to armchair in part2. As for 60° (Figure 1(c)) and 120° (Figure 1(e)), part1 and part2 have the same chirality. Here, we denote a GNR with zigzag edge in part1 and a bending angle of α as B-ZGNR-α. Besides, B-AGNR-60, B-AGNR-120 are also considered for comparison.

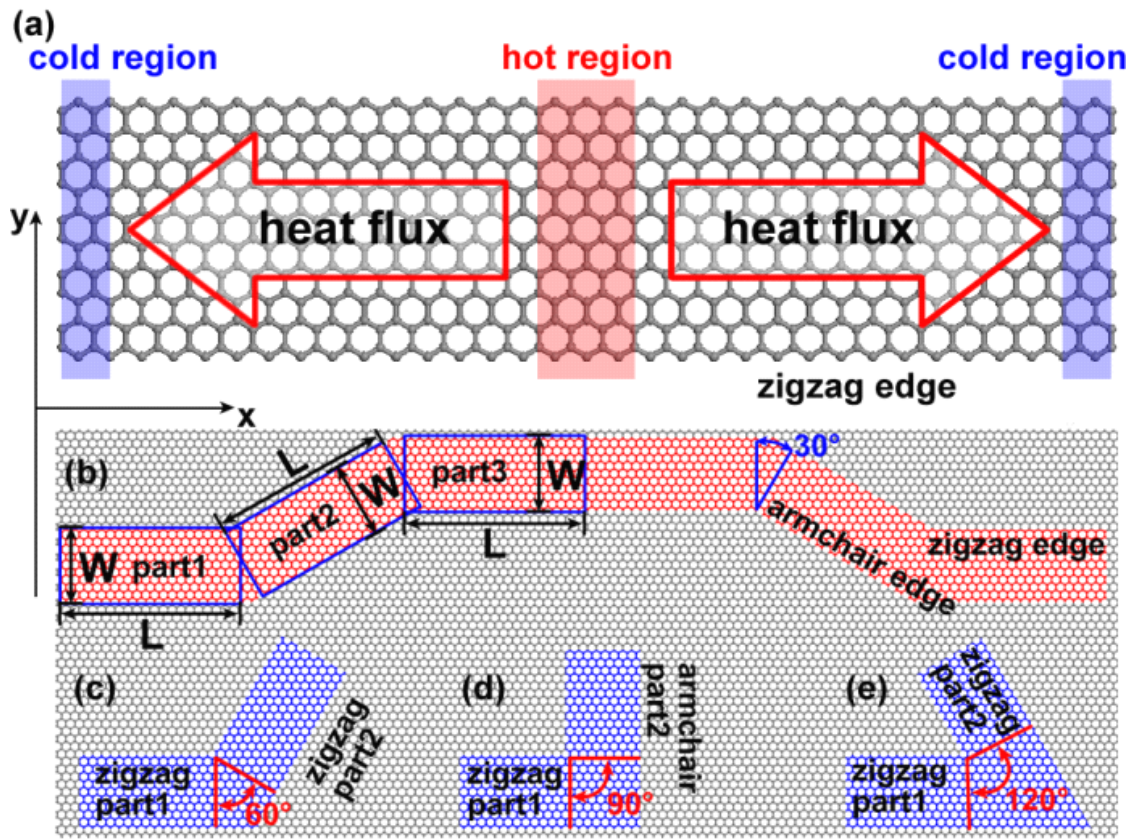

**Figure 1**. (a) Schematic picture for heat flux setting, and B-GNRs with different bending angles, (b) B-ZGNR-30, (c) B-ZGNR-60, (d) B-ZGNR-90, (e) B-ZGNR-120.

## 3. Results and discussion

### 3.1 Thermal conductivity of B-GNRs

As a test of the reliability of our approach, we first compute the thermal conductivity of ZGNR and AGNR with a length of about 11 nm, both of which have similar width about 2 nm. The result shows that the thermal conductivity of ZGNR and AGNR is 224 W/mk and 113 W/mk, which is consistent with other studies[18, 36]. Simply increasing the length of ZGNR and AGNR to 30 nm, their thermal conductivity rises to 404 W/mk and 244 W/mk. This length dependence of GNRs' thermal conductivity also agrees well with other results[18].

Next we take B-ZGNR-30, B-ZGNR-60, B-ZGNR-90 and B-ZGNR-120 (refer to Figure 1, L=5 nm, W=2 nm) as our models to study the relationship between thermal conductivity and bending angle. Figure 2 shows the temperature profile of stright ZGNR and B-ZGNR-90 through x direction. A nearly linear temperature profile is found for straight ZGNR, while a temperature jump at the corner is observed for B-ZGNR-90 due to the different directions of heat flux in different parts. The inset of Figure 2 gives the temperatue distribution of ZGNR and B-ZGNR-90, from which we see that for B-ZGNR-90 the group of atoms which are perpendicular to x direction doesn't has the same temperature. To ensure that atoms of each slab has almost the same temperature, atoms perpendicular to the geometry of B-GNR are classified into a particular slab and the average temperature of each slab are recorded (refer to the inset of Figure 2).

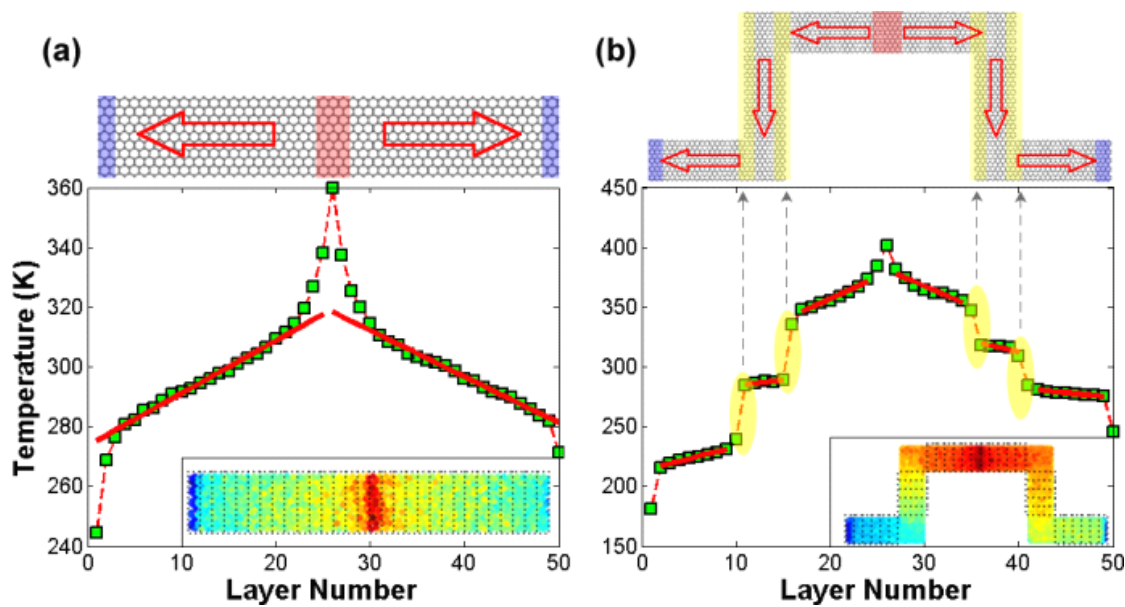

**Figure 2.** Temperature profile of (a) ZGNR and (b) B-ZGNR-90 along x direction. The inset shows the corresponding temperature distribution.

The revised temperature profiles and temperature distributions of B-ZGNRs are shown in Figure 3. The temperature gradient at each part of B-ZGNRs is computed separately, and temperature change at the corner is given as the temperature difference of adjacent part. For B-ZGNR-90, larger temperature gradient (slope of the blue lines in Figure 3) is observed in part2, which means part2 with armchair edges has relatively low thermal conductivity. This result is consistent with previous studies[18, 21] due to the lower group velocity in armchair edge direction. B-ZGNR-30 also has armchair edge in part2 which similarly leads to a larger temperature gradient. For B-ZGNR-60 and B-ZGNR-120, the temperature gradient of part1, part2 and part3 almost have the same value.

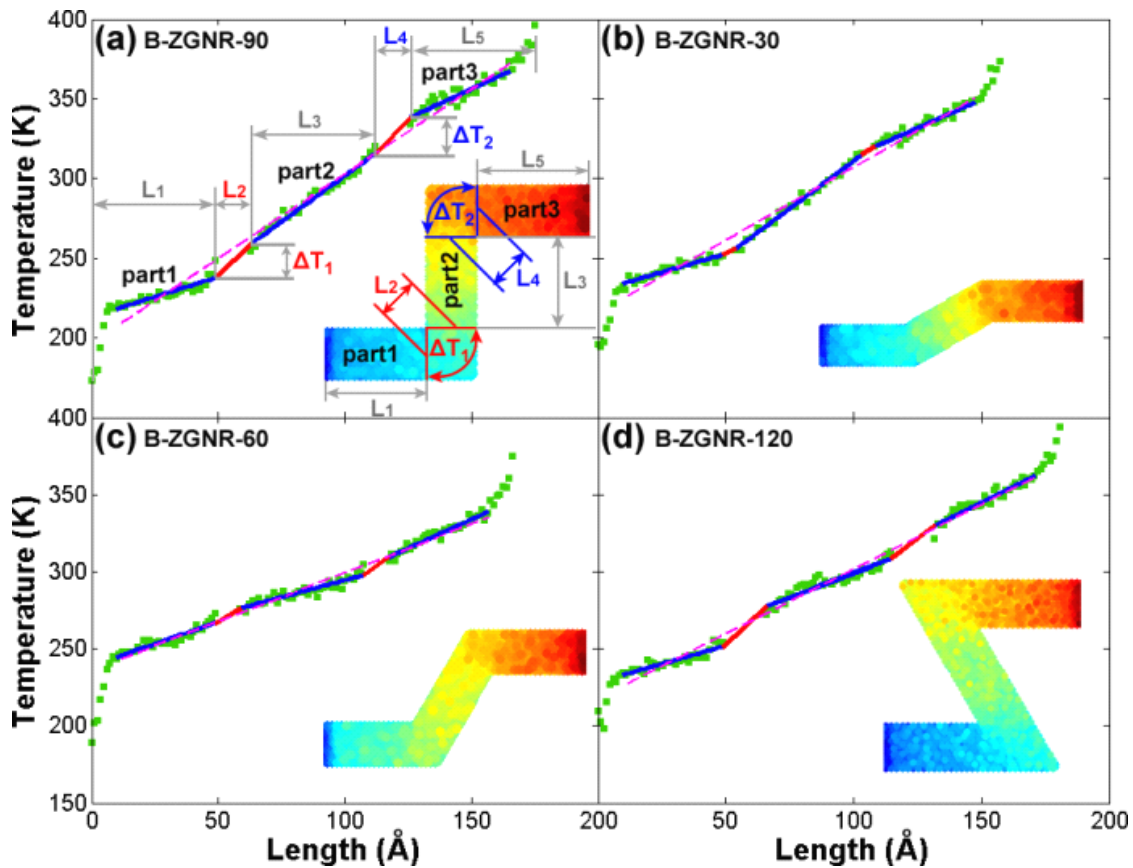

**Figure 3.** Revised temperature profiles and temperature distributions of (a) B-ZGNR-90, (b) B-ZGNR-30, (c) B-ZGNR-60, (d) B-ZGNR-120.

The thermal conductivity of B-ZGNRs with different width as a function of bending angle is shown in Figure 4. It is very interesting that the thermal conductivity of B-ZGNRs doesn't change monotonously with the bending angle. B-ZGNR-30 and B-ZGNR-90, with armchair edge in part2, have distinctly low thermal conductivity., which is reasonable considering that GNRs with zigzag edge have much higher thermal conductivity than GNRs with armchair edge. Furthermore, the results of B-ZGNR-0, B-ZGNR-60 and B-ZGNR-120 which have zigzag edge in all three parts, show a clear trend of larger bending angle with lower thermal conductivity. In adittion, for all of B-ZGNR-30, B-ZGNR-60 and B-ZGNR-90, the thermal conductivity increases monotonously with their width.

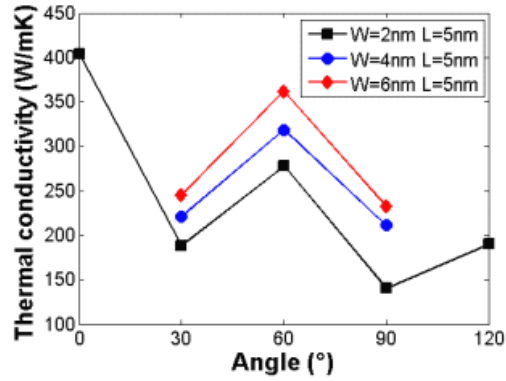

**Figure 4.** Thermal conductivity of B-ZGNRs with different width

To make a comparison with B-ZGNRs, we also compute the thermal conductivity of B-AGNR-0, B-AGNR-60 and B-AGNR-120 (W=2 nm, L=5 nm) which have armchair edge in all three parts. A summary of the results is given in Table 1. Similar to B-ZGNRs, larger bending angle of B-AGNRs also leads to lower thermal conductivity. It should be noted that the thermal conductivity of B-ZGNRs is all about 1.7 times larger than that of B-AGNRs for a specific bending angle (0°, 60° or 120°).

**Table 1.** Thermal conductivity (unit is W/mK) of B-ZGNRs and B-AGNRs

|        | 0°     | 60°    | 120°   |
|--------|--------|--------|--------|
| B-ZGNR | 404.46 | 277.85 | 190.33 |
| B-AGNR | 243.75 | 166.98 | 113.51 |

### 3.2 Kapitza conductance of B-GNRs

For B-GNRs, the bending angle will cause a Kapitza resistance at the inside corner. Taking B-ZGNR-90 for example, the temperature of edge atoms is shown in Figure 5(a). A sharp temperature jump is observed at the inside corner. By measuring the temperature jump, the Kapitza conductance can be computed through the relation[37]

$$G = \frac{J}{\Delta T} \tag{3}$$

where J is the heat flux, ΔT is the temperature jump at the inside corner. As the bending angle of B-ZGNRs changes from 30° to 120°, ΔT increases monotonously from 4.38K to 22.03K. The Kapitza conductance of B-ZGNRs with different width as a function of the bending angle is shown in Figure 5(b). The increase of bending angle leads to the decrease of Kapitza conductance while the changing of width has little effect on the Kapitza conductance.

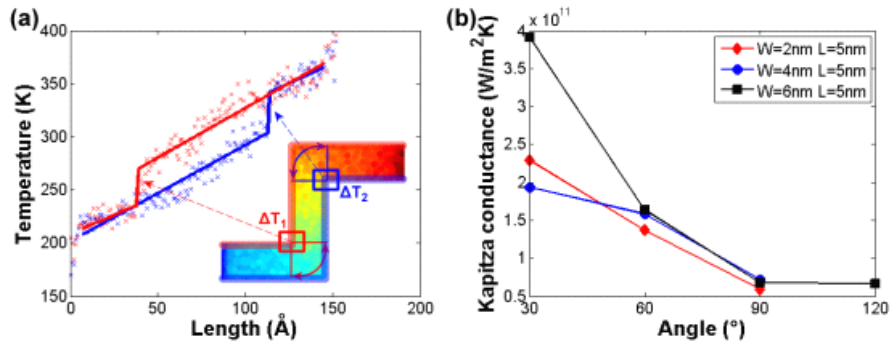

**Figure 5.** (a) Edge temperature of B-ZGNR-90. (b) Kapitza conductance of B-ZGNRs with different width

### 3.3 Modification of the corner

Modification of the inside corner of B-GNRs could improve the ability of heat transfer, considering that a majority of the thermal resistance comes from the inside corner and affects the performance of the nanoelectronic devices. B-ZGNR-90 is first considered. For convenience, $\Delta T = T_1 - T_2$ (refer to Figure 6) is used to determin the ability of heat transfer at the corner, and larger ΔT means lower ability of heat transfer. For unmodified B-ZGNR-90 (Figure 6(a)), ΔT is 52.25 K. To keep the edge smooth, two kinds of modification are applied as follows. Adding atoms at the inside corner of B-ZGNR-90 with an armchair edge (Figure 6(b)) slightly reduce ΔT to 51.99 K, while with zigzag edge(Figure 6(c)), ΔT greatly reduce to 43.08 K. That is to say, modification with zigzag edge at the inside corner could obviously improve the Kapitza conductance of B-ZGNR-90.

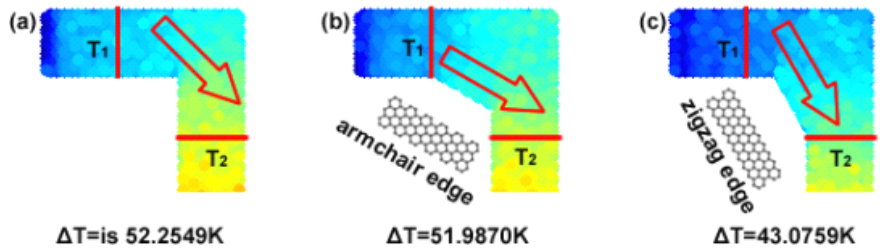

**Figure 6.** $\Delta T = T_2 - T_1$ at the corner of (a) unmodified B-ZGNR-90, (b) armchair edge modified B-ZGNR-90, (c) zigzag edge modified B-ZGNR-90.

Furthermore, we modify the inside corner of B-ZGNR-60 with armchair edge and B-AGNR-60 with zigzag edge (refer to Figure 7). ΔT increases for modified B-ZGNR-60, while ΔT decreases for modified B-AGNR-60. The chirality of modification edge distinctly affect the heat transfer at inside corner. Except the edge chirality, the size of modification also has a great impact on the ability of heat transfer. As shown in Figure 7, ΔT increases with the length of modification for B-ZGNR-60, and ΔT decreases with the length of modification for B-AGNR-60. These results show that longer modification leads to larger change of heat transport at the inside corner.

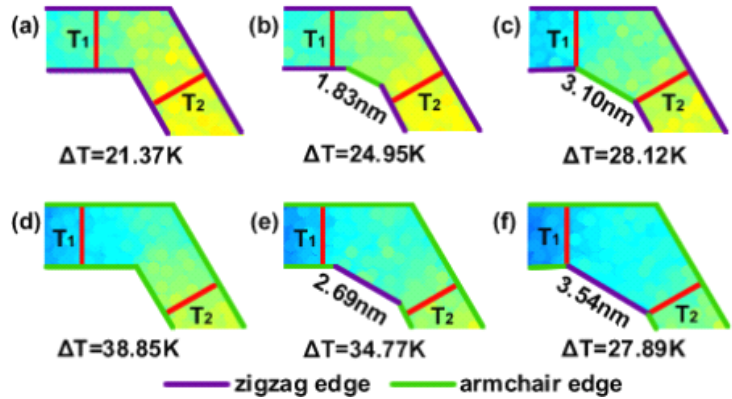

**Figure 7.** ΔT=$T_2$–$T_1$ at the corner of (a) unmodified B-ZGNR-60, (b) and (c) armchair edge modified B-ZGNR-60 with different length, (d) unmodified B-AGNR-60, (e) and (f) zigzag edge modified B-AGNR-60 with different length.

## 4. Conclusion

In summary we have investigated the thermal conductivity and Kapitza conductance of B-GNRs using the RNEMD method. The Kapitza conductance at the inside corner decreases monotonously with the bending angle. However, the thermal conductivity of B-GNRs shows a non-monotonous relationship with the bending angle. This intriguing phenomenon can be explained because the bending angle affect both chirality and the Kapitza conductance of B-GNRs. Moreover, we have modified the inside corner to guide the heat flux to another direction thus changing the ability of heat transfer at the bending place. On the basis of these information, we propose a simple relation between the thermal conductivity of the whole model and the thermal conductivity of each part, which can guide us for design of GNR-interconnected circuits.

## AUTHOR INFORMATION

Corresponding Author

*E-mail: miaoling@mail.hust.edu.cn

**ACKNOWLEDGMENT**

This work is supported by the National Natural Science Foundation of China (Grant No. ???). The authors gratefully acknowledge Dr. Xiao-jian Tan for valuable discussion on this topic.